\documentclass[lettersize,journal]{IEEEtran}
\usepackage{amsmath,amsfonts}
\usepackage{algorithmic}
\usepackage{algorithm}
\usepackage{array}
\usepackage[caption=false,font=normalsize,labelfont=sf,textfont=sf]{subfig}
\usepackage{textcomp}
\usepackage{stfloats}
\usepackage{url}
\usepackage{verbatim}
\usepackage{graphicx}
\usepackage{cite}
\usepackage{longtable}
\usepackage{makecell}
\usepackage{multicol}
\usepackage{multirow}
\begin{document}

\title{Sensing Assisted Communication for the Next Generation mmWave WLAN: System Simulation Perspective}

\author{Mao~Yang,
        Zhongjiang~Yan,
        Bo~Li,
        Qingkun~Li,
        Chenkai~Liang,
        Narengerile,
        Tony~Xiao~Han
        % <-this % stops a space
%\thanks{This paper was produced by the IEEE Publication Technology Group. They are in Piscataway, NJ.}% <-this % stops a space
%\thanks{Manuscript received April 19, 2021; revised August 16, 2021.}
\thanks{M. Yang, Z. Yan, B. Li, Q. Li, and C. Liang are with School of Electronics and Information, Northwestern Polytechnical University, Xi'an, Shaanxi, China.}
\thanks{Narengerile and T. X. Han are with Huawei Technologies Co., Ltd., China.}
}

%\Markboth{Ieee Communications Magazie}%


\maketitle

\begin{abstract}
With the proliferating of wireless demands, wireless local area network (WLAN) becomes one of the most important wireless networks. Network intelligence is promising for the next generation wireless networks, captured lots of attentions. Sensing is one efficient enabler to achieve network intelligence since utilizing sensing can obtain diverse and valuable non-communication information. Thus, integrating sensing and communications (ISAC) is a promising technology for future wireless networks. Sensing assisted communication (SAC) is an important branch of ISAC, but there are few related works focusing on the systematical and comprehensive analysis on SAC in WLAN. This article is the first work to systematically analyze SAC in the next generation WLAN from the system simulation perspective. We analyze the scenarios and advantages of SAC. Then, from system simulation perspective, several sources of performance gain brought from SAC are proposed, i.e. beam link failure, protocol overhead, and intra-physical layer protocol data unit (intra-PPDU) performance decrease, while several important influencing factors are described in detail. Performance evaluation is deeply analyzed and the performance gain of the SAC in both living room and street canyon scenarios are verified by system simulation. Finally, we provide our insights on the future directions of SAC for the next generation WLAN.
\end{abstract}


\section{Introduction}
\IEEEPARstart{W}{ireless} networking has experienced and is undergoing rapid development and evolution, and has become one of the most important infrastructures affecting people's lives. Wireless local area network (WLAN) and cellular network together become the most important types of broadband wireless network storing, transmitting, and computing network information. At present, 5G and WiFi 6 are being rapidly deployed around the world, while academia and industry are focusing on the research of next generation wireless networks. In the evolution of next generation wireless networks, simply improving the extreme performance of the network such as peak transmission rate is difficult to keep up with the increasingly diverse services and user requirements. Therefore, making wireless networks intelligent is an important and promising objective for the next era.

A basic requirement for network intelligence is to achieve abundant and diverse information to further learn and make decisions. However, it is not enough to rely only on communications information, because the networked people and devices are in the surroundings, and communication information is quite difficult to describe the changes and features of the surroundings such as location and moving information, channel information, \emph{and etc}.. We need to use non-communication methods to obtain surrounding information and, consequently, integrating sensing and communications (ISAC) is considered as one of the most appropriate means \cite{ISAC_Survey,ISAC_Survey2}. At present, the ISAC has been considered as one key technology of 6G \cite{ISAC_Cellular}, while IEEE has established a standardization task group of WLAN sensing: IEEE 802.11bf \cite{11bf_Survey2}.

We are witnessing the rapid growth of WLAN. Based on Cisco's report\cite{Cisco}, the number of WiFi 6 hotspots will increase 13 times from 2020 to 2023, and 51 percent of global IP traffic will be offloaded by Wi-Fi. Sensing assisted communication (SAC), utilizing the sensing results to assist and improve the performance of communication, is an important branch of ISAC. It is more practical and necessary to realize SAC in the next generation WLAN because of short distance, open to deploy and integrating, and rapid evolution \cite{Sensing_WiFi}. Especially, the millimeter wave WLAN \cite{11ay_survey} operating on the above-45GHz band can benefit from SAC since the physical layer (PHY) and the media access control layer (MAC) towards directional transmission are more sensitive to the location information and the channel information. WiFi 7 standardized by IEEE 802.11be and WLAN sensing standardized by IEEE 802.11bf will be released in 2024, and IEEE 802.11 will soon establish a study group for WiFi 8 in 2022. Therefore, this article focuses on SAC in the next generation WLAN.

The deployment of the SAC into the next generation WLAN will benefit the network performance. For example, the sensing of user behavior from gesture or other features is conducive to guarantee the quality of user experiences (QoE). The sensing of location and mobility is very valuable for dynamically beam switching and seamless handoff. The sensing of channel information is quite helpful to the improvement of multiple-input multiple-output (MIMO) performance and user grouping. However, most related works pay attention to the PHY technologies, there are few related works systematically analyzing the gain, influencing factors, and the performance evaluations on the SAC in the next WLAN.

To the best of our knowledge, this is the first work to systematically study and analyze SAC in the next generation WLAN from the system simulation perspective. This article analyzes the scenarios and advantages of SAC in the next generation WLAN. After that, from system simulation perspective, we analyze the source of network performance gain brought by the SAC, and discuss a series of key influencing factors in detail, and evaluate the performance gain through system level simulation. Finally, we provide some future directions of SAC for the next generation wireless networks.

\section{Scenarios and Advantages of SAC of the Next Generation WLAN}

\begin{figure}[!t]
\centering
\includegraphics[width=3in]{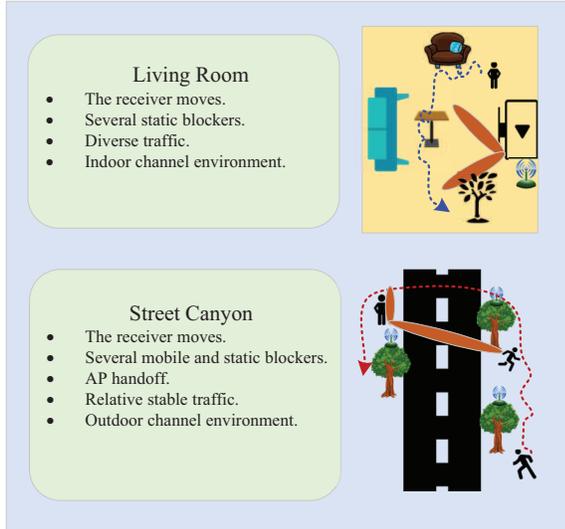}
\caption{Example Scenarios of SAC in the Next Generation WLAN.}
\label{fig_Advantages}
\end{figure}

We divide network information into communications space (CS) and sensing space (SS). CS denoted as $\mathbf{C}$ means the communication related information such as transmission power, signal to interference and noise ratio (SINR), and bandwidth. SS denoted as $\mathbf{S}$ means the sensing related information such as location and mobility, channel information, and user preference. The network tries to guarantee the network performance, named performance space $\mathbf{P}$ depicting the network performance and the user experiences. As long as the communication information is obtained, we can obtain the corresponding network performance through relevant methods. Only communication information is not enough for us to grasp the whole picture of information transmission, so the obtained network performance will be uncertain, which is denoted as $H\left( {\bf{P}} \right)$, where $H\left( {} \right)$ is the entropy function. Integrating sensing into communications means both the communication information and the sensing information can be simultaneously obtained. Consequently, the uncertainty of network performance will be reduced:

\begin{equation} \label{Eq_uncertainty}
H\left( {\bf{P}} \right) \ge H\left( {{\bf{P}}|{\bf{C}}} \right) \ge H\left( {{\bf{P}}|{\bf{C}},{\bf{S}}} \right).\
\end{equation}
Eq. (\ref{Eq_uncertainty}) means the additionally obtained sensing information can increase the certainty of network performance. It is worth noting that less uncertainty does not mean better network performance. If the sensing information is a positive priori, the network performance can be improved. For example, very accurate sensed user location information brings much accurate $\left\langle {tx,rx} \right\rangle $  beam pair. On the contrary, if the sensing information is a negative priori, maybe the network performance will deteriorate. For example, the location sensing with large error will lead to serious deviation from the optimality of the beam pair. Therefore, better sensing performance is very important for SAC.

Fig. \ref{fig_Advantages} shows the example scenarios of SAC in the next generation WLAN including the living room and the street canyon, which are also the target simulation scenarios in the following sections. We illustrate advantages brought from SAC according to the example scenarios shown in Fig. \ref{fig_Advantages}.

QoE is very subjective and highly correlated with user preferences. It is difficult to guarantee the QoE simply based on communication information, because communication information is insensitive to subjective characteristics. If the subjective characteristics such as user behaviors can be sensed, the WLAN will provide a series of methods to improve the QoE. Note that WLAN covers short-range local area, which makes it more easier to obtain user behavior information based on gestures, locations and other information \cite{Sensing_Gesture}. As the living room scenario shown in the Fig. \ref{fig_Advantages}, the access point (AP) or other sensing device senses that the user moves to the sofa (location) and stays still (mobility), and the mobile phone keeps upright (gesture) at 9-10pm (time duration). The AP or the controller determines that the short video service especially the sports news will probably be the user preference at this time and, consequently, several possible methods can be utilized to guarantee the QoE, e.g., intelligent information pushing, caching, and content centric routing. Also, the AP may provide appropriate QoS guarantee schemes, e.g., uplink scheduling in WiFi 6, restricted target wake-up time (r-TWT) and traffic identifier (TID) to link mapping in WiFi 7 \cite{11be_survey2,11be_survey3}.

Location and mobility features are quite important for wireless networks. Different locations including the transmitter, the receiver, and the blocker lead to quite different channel environment, interferences, and collision status, directly affecting the throughput and the reliability. The mobility feature affects the stochastic process of communication information varying with time. Especially for the millimeter wave WLAN, the directional transmission is quite sensitive to the location and mobility. As both scenarios shown in the Fig. \ref{fig_Advantages}, the AP senses the mobility of the user, then it dynamically uses the optimal $\left\langle {tx,rx} \right\rangle $  beam pair and consistently guarantees the communication quality. Yuan et al. \cite{Radar} propose a quasi-zero overhead for beamtrcking by using Bayesian predication in the vehicular network. Polese et al. \cite{Sensing_CSI} adopt a machine learning based data driven scheme for beam switching without complicated training procedure. Liu et al \cite{ISAC_Channel} proposes a learning based channel state prediction scheme for beamforming.

\section{Performance KPI, Gain and Impact Factors of Sensing Assisted Communication (SAC)}

\begin{figure*}[!t]
\centering
\includegraphics[width=7in]{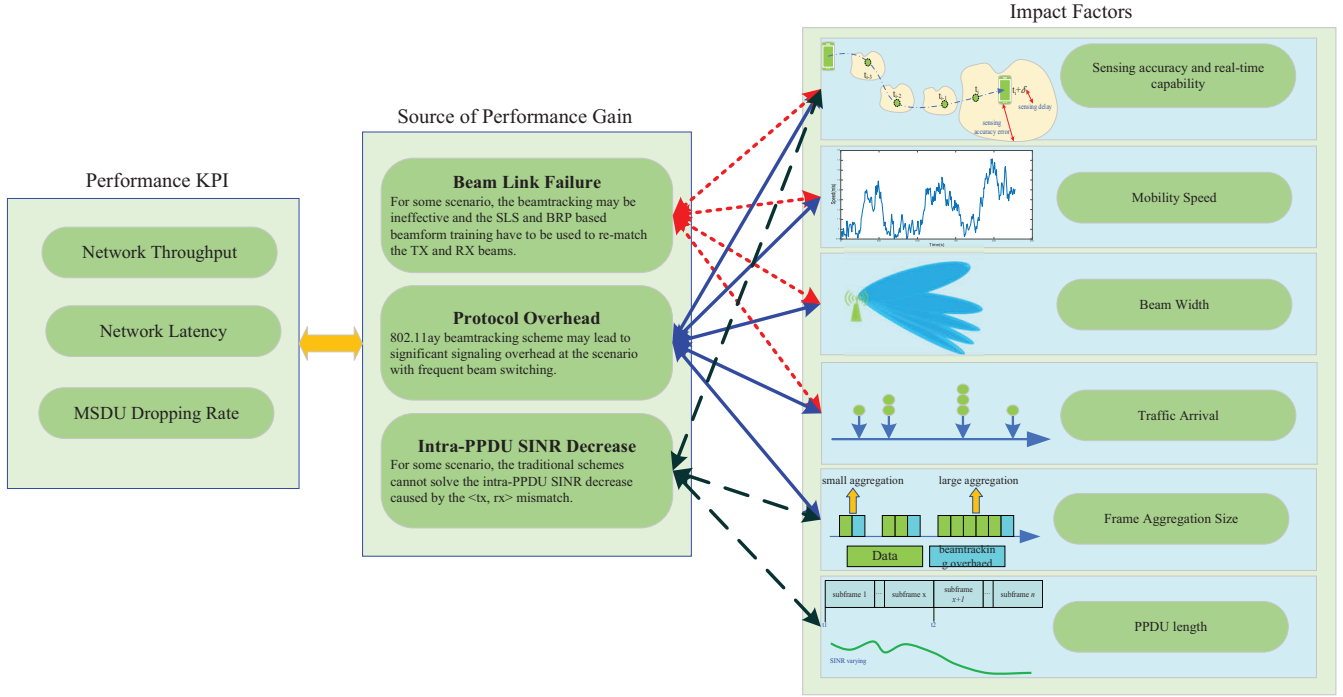}
\caption{KPI, gain and influencing factors of SAC for the next generation Wi-Fi.}
\label{fig_Gain}
\end{figure*}

To deeply study and evaluate the potential benefits obtained from sensing, this section analyzes the performance key performance indicator (KPI), source of performance gain, and important impact factors from the system simulation perspective.

\subsection{Performance KPI of SAC}

To evaluate the performance gain of the SAC, some performance KPI are recommended.
\begin{itemize}
  \item Network throughput. Throughput is defined as successfully transmitted bits in a unit time by one network or one communication link. It is a comprehensive metric of system performance. Especially, the saturated throughput can reflect the maximum capacity of the network.
  \item Network latency. Latency is defined as the average time each packet takes from entering the MAC layer to reaching the destination node's MAC layer successfully. The latency can comprehensively reflect how long a packet has experienced queuing, channel access, transmission, retransmit, and acknowledgement.
  \item MAC service data unit (MSDU) dropping rate. MSDU dropping rate is defined as the percentage of packets that are dropped by the MAC layer queue because they exceed the maximum tolerance time or the maximum number of retransmissions.
\end{itemize}

\subsection{Source of Performance Gain}

As Fig. \ref{fig_Gain} depicts, there are three main sources of performance gain for Wi-Fi brought from SAC: beam link failure, protocol overhead, and intra-PHY protocol data unit (intra-PPDU) SINR decrease.

\textbf{\romannumeral1 ) Beam link failure.}

For some scenarios, e.g. the node is moving and its traffic is delay sensitive with low traffic rate, the beamtracking may become ineffective. The reason is that if the traffic of the node is low, there will be a long traffic interval between the current arrival packets and the previous packets. During this long time interval, the node may move a long distance. Thus, the original trained $\left\langle {tx,rx} \right\rangle $ beam pair may be far from the optimal one. This causes that the preamble of the PPDU may not be received correctly (it means the signal synchronization failed) and, consequently, the beamtracking process becomes ineffective. This issue is called beam link failure problem. Therefore, without the SAC, the sector-level sweep (SLS) and/or beam refinement protocol (BRP) have to be used to re-train the TX and RX beams, causing a very long time of communication interruption and directly deteriorating the QoS and QoE. On the contrary, the SAC helps the nodes to select better beam pairs based on the sensing of channel information and the location/mobility, avoiding communication interruption.

\textbf{\romannumeral2 ) Protocol overhead.}

Consider the scenario with beam switching caused by the node movement, frequent blocking, \emph{and etc}., beamtracking is supposed to be a method to dynamically improve the beam switching performance for the mmWave Wi-Fi. Every time beamtracking is used, it will bring both the signaling overhead and the air interface overhead between the transmitter and the receiver of one beam link because beamtracking adds the beam training process and the corresponding fields at the tail of the data PPDU. When the beam switching becomes increasingly frequent, more beamtracking processes and corresponding fields are needed, resulting in an increasing overhead of beamtracking without the SAC. Furthermore, the PHY rate of millimeter wave Wi-Fi is usually much larger than that of the sub-7 GHz Wi-Fi, which further increases the overhead of beamtracking because the air interface duration of beam training accounts for a larger proportion of the PPDU duration. This significantly reduces the throughput of the system when traffic rate is high. On the contrary, the ISAC enables the nodes to sense the location of the transmitters, receivers, and the blockers to dynamically switch the beam pair based on the sensing results without or with little overhead.

\textbf{\romannumeral3 ) Intra-PPDU SINR decrease.}

In Wi-Fi, the PPDU length may be long for several $ms$ level because of frame aggregation, lower MCS, small number of spatial streams, narrow bandwidth, \emph{and etc}.. Without the SAC, the SINR may significantly decrease during the PPDU transmission period (called intra-PPDU) because of $\left\langle {tx,rx} \right\rangle $ beam pair mismatch caused by the node mobility, blocking, narrow beam width, \emph{and etc}.. Thus, the packet loss rate is increased intra-PPDU, resulting in throughput decrease. Unfortunately, there are no effective means to solve the intra-PPDU SINR decrease problem, when SAC is not considered. An intuitive solution is to reduce the length of PPDU, but this leads to low MAC efficiency and also reduces the throughput. With the SAC, the transmitter and the receiver can dynamically adjust the transmission parameters according to the sensing results to improve the intra-PPDU SINR and packet loss rate.

\subsection{Impact Factors of the SAC}

Fig. \ref{fig_Gain} shows six impact factors: sensing accuracy and real-time capability, mobility speed, beam width, traffic arrival, frame aggregation size, and PPDU length. It also depicts the relationship between the source of performance gain brought from the SAC and these influencing factors.

\textbf{\romannumeral1 ) Sensing accuracy and real-time capability.}

Sensing accuracy and real-time capability directly affect the SAC performance. Higher sensing accuracy helps the nodes to determine the optimal communication schemes and parameters, while lower sensing accuracy causes the determined communication parameters to deviate from the optimal value. Similarly, better real-time capability makes the sensed results closer to the current status.

\textbf{\romannumeral2 ) Mobility speed.}

With the nodes including the transmitter, the receiver and the blocker move, the performance probably changes. Thus, mobility speed is an obvious influencing factor. The faster the moving speed, the more dramatic changes will occur in channel state and the $\left\langle {tx,rx} \right\rangle $ beam pair, which will worsen the performance without the SAC. The SAC-based scheme can dynamically adjust the communication parameters through sensing the channel information and location/mobility information.

\textbf{\romannumeral3 ) Beam width.}

Beam width is an important parameter affecting the transmission quality of the millimeter wave Wi-Fi. A narrower beam width can better suppress side lobe leaks, but it will make the communications more sensitive to the node mobility. In other words, a very narrow beam width causes frequent switching of $\left\langle {tx,rx} \right\rangle $ beam pair, which is a challenge to the beamtracking scheme. Furthermore, the narrower the beam, the more likely it is to cause the blockers to interrupt communication.

\textbf{\romannumeral4 ) Traffic arrival.}

The traffic characteristic influences the performance gain brought from the SAC. For example, when the traffic rate is low, the beam link failure problem predominates and the beamtracking mechanism becomes ineffective, resulting in large delay and MAC service data unit (MSDU) loss rate without the SAC. When the traffic rate is high, the ISAC-based scheme significantly outperforms the non-sensing scheme in saturated throughput because of very tiny protocol overhead.

\textbf{\romannumeral5 ) Frame aggregation size.}

Frame aggregation size deeply influences the MAC efficiency. The smaller the frame aggregation size, the lower the MAC efficiency and the higher protocol overhead of beamtracking. Conversely, the larger the frame aggregation size, the higher the MAC efficiency and the lower protocol overhead of beamtracking. However, too large frame aggregation size leads to very long PPDU length, which will cause the intra-PPDU SINR decrease problem.

\textbf{\romannumeral6 ) PPDU length.}
The PPDU length reflects the duration of the PHY frame at the air interface. A longer PPDU length will cause the intra-PPDU SINR decrease problem and, consequently, lead to lower throughput and higher packet loss rate.

\section{Performance Evaluation}
\subsubsection{Simulation Scenario and Settings}

Fig. \ref{fig_Advantages} shows the scenarios of SAC in the next generation Wi-Fi including the living room and the street canyon, which are also the target simulation scenarios in the following sections. We illustrate advantages brought from SAC according to the example scenarios shown in Fig. \ref{fig_Advantages}. The living room represents the indoor scenario, while the street canyon represents the outdoor scenario. There are one AP and one target STA in both scenarios, and the STA keeps moving in the scenarios. In the living room scenario, the beam will be reflected by the ceiling, floor, wall and furniture. The direct path will be blocked by some furniture and human body. In the street canyon scenario, the beam will be reflected or blocked by the ground, trees, buildings on the street and pedestrians on the street. The AP and the target STA communicate by using mmWave beam link. The moving speed and the beam width can be changed. The detailed parameter settings will be described in the next subsection.

\subsubsection{Simulation Results}
To deeply analyze the performance of SAC enabled next generation Wi-Fi, we deploy a series of simulations. We focus on two scenarios: living room and street canyon as shown in Fig. \ref{fig_Advantages} suggested and described in the IEEE 802.11ay document. We established an integrated system level (protocol and signaling of the MAC and the upper layers) and link level (channel model and PHY abstraction) simulation platform towards millimeter wave Wi-Fi based on network simulator 3 (NS3), and integrate sensing into the simulation platform to evaluate the performance gain of the SAC. We use the ray tracing method to model the channel based on the layout of each scenario and the antenna pattern.

\textbf{\romannumeral1 ) Simulation Results of the Living Room Scenario.}

There are one AP and one STA in the living room, and the STA randomly moves in the room. It is worth noting that the change of relative position between nodes needs to be depicted in system simulation, whether the moving node is the transmitter, receiver or the blocker, the final effect is equivalent. There are $18$ beam sectors in the ${360^o}$ range. Both uplink and downlink possess traffic to transmit and have the same traffic rate. Unless otherwise specified, the frame aggregation size is 1. The modulation and coding scheme (MCS) index is 14 of IEEE 802.11ay. There are three schemes:

\begin{itemize}
  \item 802.11ay with beamtracking indicates that the system strictly works according to the 802.11ay standard with the beamtracking function.
  \item The ISAC scheme indicates the AP and the STA dynamically switch their $\left\langle {tx,rx} \right\rangle $ beam pair according to the location sensing results. The sensing accuracy error varies, which means the sensed locations are randomly distributed around the real location and the radius of the sensing accuracy error.
  \item Theoretical optimum means the AP and the STA can always select the optimal $\left\langle {tx,rx} \right\rangle $ beam pair based on the real location without any overhead.
\end{itemize}

\begin{figure}[!t]
\centering
\includegraphics[width=3.5in]{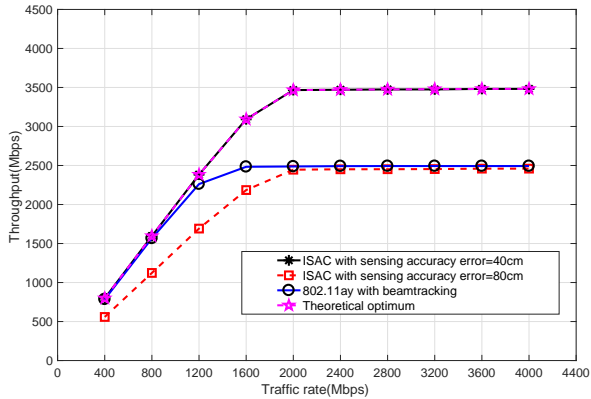}
\caption{Throughput performance comparison with traffic rate varying in the living room scenario.}
\label{6-17}
\end{figure}

\begin{figure}[!t]
\centering
\includegraphics[width=3.5in]{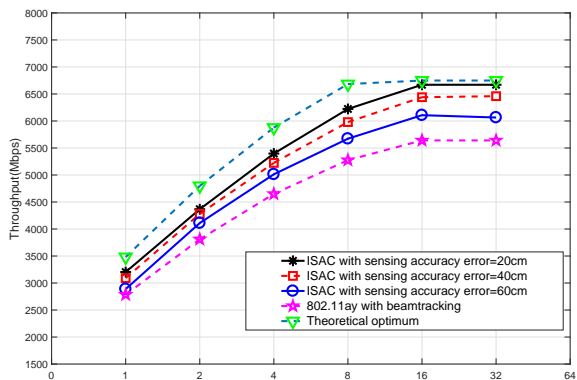}
\caption{Throughput performance comparison with frame aggregation size in the living room scenario.}
\label{6-35}
\end{figure}

\begin{figure}[!t]
\centering
\includegraphics[width=3.5in]{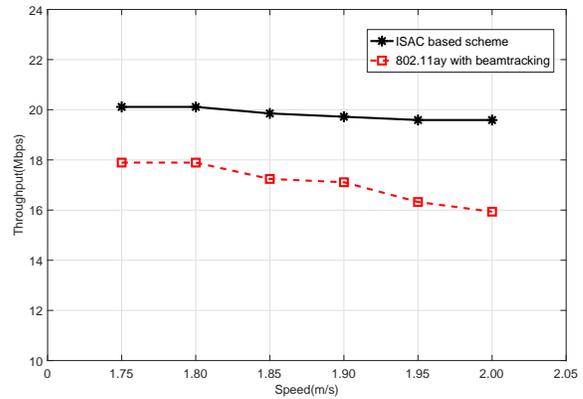}
\caption{Throughput performance comparison with moving speed varying in the living room scenario.}
\label{6-38}
\end{figure}

\begin{figure}[!t]
\centering
\includegraphics[width=3.5in]{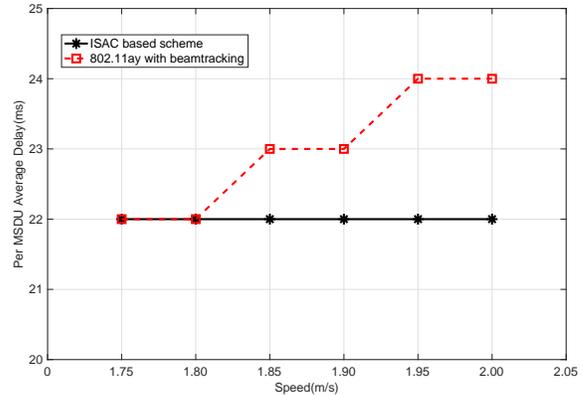}
\caption{Per packet delay with moving speed varying in the living room scenario.}
\label{6-39}
\end{figure}

\begin{figure}[!t]
\centering
\includegraphics[width=3.5in]{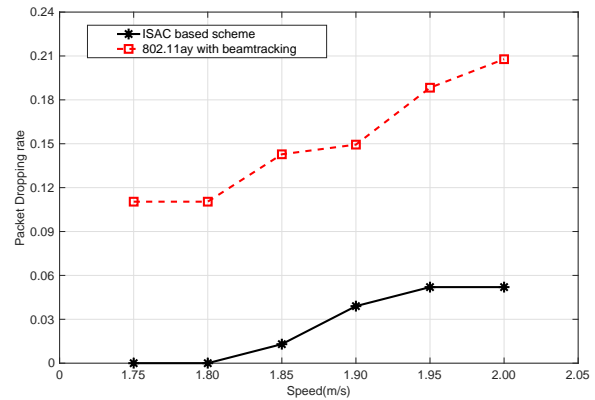}
\caption{Packet dropping rate with moving speed varying in the living room scenario.}
\label{6-40}
\end{figure}

Fig. \ref{6-17} shows the throughput performance comparison with traffic rate varying in the living room scenario. The value in the X-axis indicates the traffic rate for both uplink and downlink. It can be observed that the throughput achieved by all schemes increases first and then stabilizes with the increase of traffic rate. This is because the network has reached saturation from unsaturated state. It is worth noting that even if the sensing error reaches $40cm$ that is not a small value, the throughput of the ISAC scheme is still very close to the theoretical performance. The theoretical performance means the sender and the receiver can always use the theoretically optimal beam pair and, consequently, can obtain the optimal throughput. The throughput of the 802.11ay with beamtracking scheme is lower. This is because frequently beamtracking leads to considerable signaling overhead. Moreover, when the sensing error is very large, for example $80cm$ in Fig. \ref{6-17}, the performance of the ISAC scheme deteriorates since large sensing error leads to serious mismatch beam pair. Thus, on the one hand, the source of the gain of this simulation is the signaling overhead. On the other hand, the sensing accuracy is a parameter that significantly affects system performance and should therefore be as accurate as possible.

Fig. \ref{6-35} shows the throughput performance comparison with frame aggregation size varying in the living room scenario. In this scenario, the traffic rate is $4,000Mbps$ for both uplink and downlink. It can be observed that the throughput achieved by all schemes increases first and then stabilizes with the increase of frame aggregation size. This is because frame aggregation can enhance the MAC efficiency and the network has reached saturation. It is worth noting that the throughput of the ISAC scheme is close to the theoretical performance. Even if the sensing error reaches $60cm$ that is a relatively large value in the living room scenario, the throughput of the ISAC scheme still outperforms the 802.11ay with beamtracking scheme because of significantly low protocol overhead.

Fig. \ref{6-38} shows the throughput performance comparison with moving speed varying in the living room scenario. In this simulation, the traffic rate is unsaturated, $10Mbps$ for both uplink and downlink. It can be seen that with the increase of the moving speed, the throughput of the ISAC scheme is stable and close to $20Mbps$. This is because the ISAC scheme can serve almost all the packets in time. Considering that the traffic rate is $10Mbps$, the total traffic rate of the uplink and the downlink is $20Mbps$. It is interesting although the traffic rate is unsaturated, the throughput of the 802.11ay with beamtracking scheme decreases with the increase of the moving speed. The reason is that unsaturated traffic leads to large packet arrival interval and correspondingly large move distance, which causes beam link failure. When beam link failure occurs, the beamtracking becomes ineffective because preamble cannot be successfully received. Then, the AP and the STA have to wait for the next beacon interval to re-train the beamform link. This process takes a long time. However, the sender continuously tries to retransmit data. When the maximum retransmission time is reached, the packet (MSDU) will be dropped from the MAC queue.

Fig. \ref{6-39} shows the per packet average delay performance comparison with moving speed varying in the living room scenario. It can be seen that with the increase of the moving speed. The delay of the ISAC scheme is quite stable. In the unsaturated scenario, almost every packet can be transmitted in time because the sender and the receiver always utilize the approximating optimal beam pair. However, the per packet average delay of the 802.11ay with beamtracking scheme keeps increasing with the increase of the moving speed. The beam link failure problem occurs results longer waiting time to re-train the beamform link. Low latency traffic or named real-time application is one important requirement in the future, the ISAC scheme can significantly guarantee the delay performance.

Fig. \ref{6-40} shows the packet dropping rate performance comparison with moving speed varying in the living room scenario. The per packet dropping rate of the ISAC scheme increases slowly with the increase of the moving speed because of the sensing error. In this scenario, the traffic rate is small, almost every packet can be transmitted in time. However, the per packet dropping rate of the 802.11ay with beamtracking scheme is significantly larger than that of the ISAC scheme. Because this unsaturated scenario leads to beam link failure problem, which causes longer waiting time to re-train the beamform link. During the waiting time, the 802.11ay with beamtracking scheme still retransmits the packets and, finally, the maximum retransmission time is reached and the packets are dropped from the queue.

\textbf{\romannumeral2 ) Simulation Results of Street Canyon Scenario.}

There are one AP and one STA in the street canyon, and the STA walks along the street. It is worth noting that the change of relative position between nodes needs to be depicted in system simulation, whether the moving node is the transmitter, receiver or the blocker, the final effect is equivalent. There are $18$ beam sectors in the ${360^o}$ range. Both uplink and downlink possess traffic to transmit and have the same traffic rate. Unless otherwise specified, the frame aggregation size is 1. The modulation and coding scheme (MCS) index is 14 of IEEE 802.11ay. There are three schemes:

\begin{itemize}
  \item 802.11ay with beamtracking indicates that the system strictly works according to the 802.11ay standard with the beamtracking function.
  \item The ISAC scheme indicates the AP and the STA dynamically switch their $\left\langle {tx,rx} \right\rangle $ beam pair according to the location sensing results. The sensing accuracy error varies, which means the sensed locations are randomly distributed around the real location and the radius of the sensing accuracy error.
  \item Theoretical optimum means the AP and the STA can always select the optimal $\left\langle {tx,rx} \right\rangle $ beam pair based on the real location without any overhead.
\end{itemize}

\begin{figure}[!t]
\centering
\includegraphics[width=3.5in]{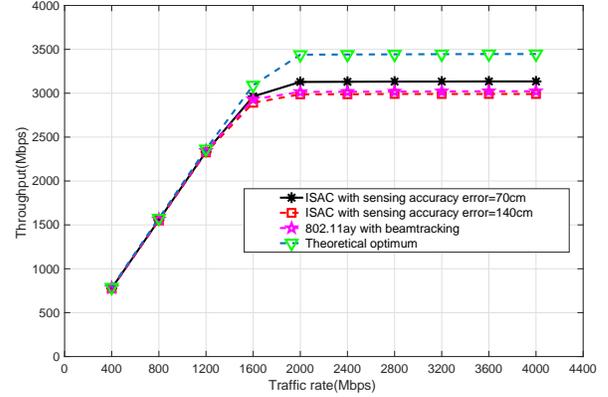}
\caption{Throughput performance comparison with traffic rate varying in the street canyon scenario.}
\label{7-7}
\end{figure}

\begin{figure}[!t]
\centering
\includegraphics[width=3.5in]{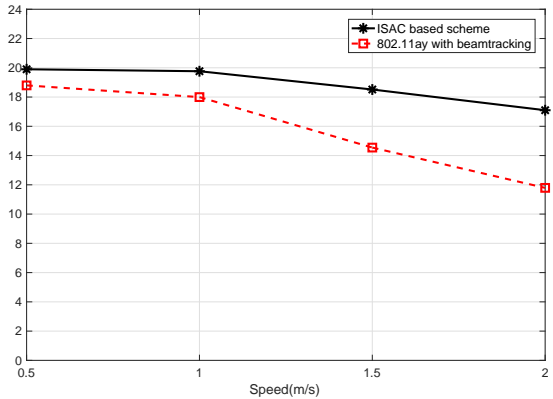}
\caption{Throughput performance comparison with moving speed varying in the street canyon scenario.}
\label{7-16}
\end{figure}

\begin{figure}[!t]
\centering
\includegraphics[width=3.5in]{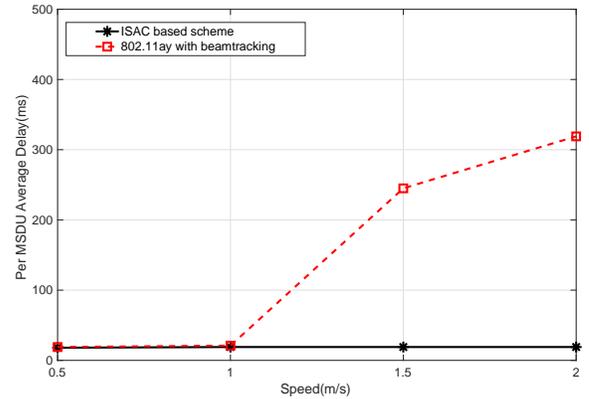}
\caption{Per packet delay with moving speed varying in the street canyon scenario.}
\label{7-17}
\end{figure}

\begin{figure}[!t]
\centering
\includegraphics[width=3.5in]{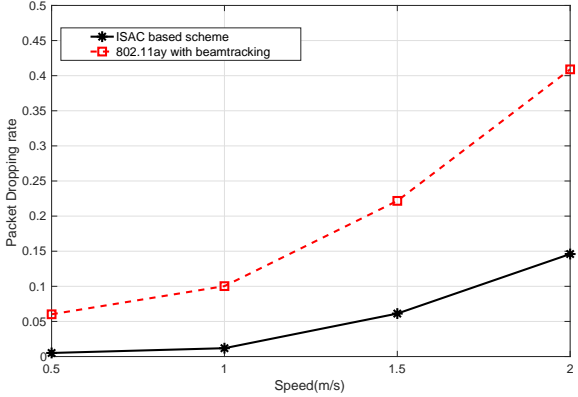}
\caption{Packet dropping rate with moving speed varying in the street canyon scenario.}
\label{7-18}
\end{figure}

\begin{figure}[!t]
\centering
\includegraphics[width=3.5in]{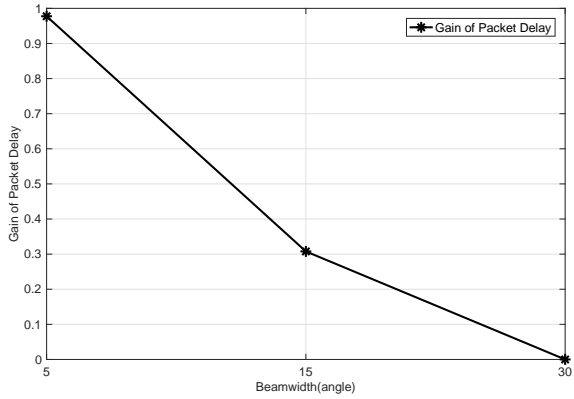}
\caption{Gain of the per packet average delay performance comparison with the beam width varying in the street canyon scenario}
\label{7-25}
\end{figure}

\begin{table*}[!ht]
\renewcommand{\arraystretch}{1.3}
\caption{Performance analysis and insights of SAC}
\label{tab:1}
\centering
% Some packages, such as MDW tools, offer better commands for making tables
% than the plain LaTeX2e tabular which is used here.
\begin{tabular}{|c|c|m{8cm}|}
\hline
\bfseries Influencing Factors & \bfseries Importance & \bfseries  Performance Analysis based on System Simulations\\
\hline
Sensing accuracy and real-time capability & High & Higher sensing accuracy and real-time capability achieve large performance gain. Lower sensing accuracy and real-time capability result in smaller performance gain, even performance loss if the sensing accuracy is quite low.\\
\hline
Traffic arrival & High & Low traffic rate: beam link failure problem occurs and the beamtracking becomes ineffective, resulting in large delay and MSDU loss rate. However, the SAC-based scheme helps to guarantee the performance. High traffic rate: sensing scheme significantly outperforms the non-SAC scheme in saturated throughput because of very low protocol overhead.\\
\hline
Mobility speed & High & As the moving speed increases, the throughput of the SAC-based scheme remains stable, but that of the non-SAC scheme continues to decline. The delay of the SAC-based scheme remains stable, but that of the non-SAC scheme increases significantly. The MSDU loss rate of the SAC-based scheme keeps small, but that of the non-SAC scheme increases significantly.\\
\hline
Beam width & High & As the beam width is getting narrow, the throughput of the SAC-based scheme remains stable, but that of the non-SAC scheme continues declining. The delay of the SAC-based scheme remains stable, but that of the non-SAC scheme increases significantly.
The MSDU loss rate of the SAC-based scheme keeps small, but that of the non-SAC scheme increases sharply.\\
\hline
Aggregation size & Middle & As the aggregation size is small, the beamtracking overhead is significant. As it is large, the beamtracking overhead is unapparent but the intra-PPDU SINR decrease problem frequently occur.\\
\hline
PPDU length & Middle & As the PPDU length increases, the air interface packet loss rate of the non-SAC scheme becomes larger because of intra-PPDU SINR decreases, resulting in low throughput. The SAC-based scheme may introduce the intra-PPDU beam switch based on the sensing results during the PPDU duration, which guarantees the throughput.\\
\hline
\end{tabular}
\end{table*}

Fig. \ref{7-7} shows the throughput performance comparison with traffic rate varying in the street canyon scenario. The value in the X-axis indicates the traffic rate for both uplink and downlink. It can be observed that the throughput achieved by all schemes increases first and then stabilizes with the increase of traffic rate. This is because the network has reached saturation from unsaturated state. The throughput of the ISAC scheme outperforms the 802.11ay with beamtracking scheme. This is because frequently beamtracking leads to considerable signaling overhead. When the sensing error becomes larger, the performance gain will be reduced.

Fig. \ref{7-16} shows the throughput performance comparison with moving speed varying in the street canyon scenario. In this simulation, the traffic rate is unsaturated, $10Mbps$. It can be seen that with the increase of the moving speed, the throughput of the ISAC scheme is stable and close to $20Mbps$. This is because the ISAC scheme can serve almost all the packets in time. Considering that the traffic rate is $10Mbps$, the total traffic rate of the uplink and the downlink is $20Mbps$. It is interesting although the traffic rate is unsaturated, the throughput of the 802.11ay with beamtracking scheme decreases with the increase of the moving speed. The reason is that unsaturated traffic leads to large packet arrival interval and correspondingly large move distance, which causes beam link failure. When beam link failure occurs, the beamtracking becomes ineffective because preamble cannot be successfully received. Then, the AP and the STA have to wait for the next beacon interval to re-train the beamform link. This process takes a long time. However, the sender continuously tries to retransmit data. When the maximum retransmission time is reached, the packet (MSDU) will be dropped from the MAC queue.

Fig. \ref{7-17} shows the per packet average delay performance comparison with moving speed varying in the street canyon scenario. It can be seen that with the increase of the moving speed. The delay of the ISAC scheme is quite stable and very small. In the unsaturated scenario, almost every packet can be transmitted in time because the sender and the receiver always utilize the approximating optimal beam pair. However, the per packet average delay of the 802.11ay with beamtracking scheme sharply increases with the increase of the moving speed. The beam link failure problem occurs results longer waiting time to re-train the beamform link.

Fig. \ref{7-18} shows the packet dropping rate performance comparison with moving speed varying in the street canyon scenario. The per packet dropping rate of the ISAC scheme increases slowly with the increase of the moving speed because of the sensing error. In this scenario, the traffic rate is small, almost every packet can be transmitted in time. However, the per packet dropping rate of the 802.11ay with beamtracking scheme is significantly larger than that of the ISAC scheme. In this figure, the packet dropping rate of the 802.11ay with beamtracking scheme even exceeds 40 percent. Because this unsaturated scenario leads to beam link failure problem, which causes longer waiting time to re-train the beamform link. During the waiting time, the 802.11ay with beamtracking scheme still retransmits the packets and, finally, the maximum retransmission time is reached and the packets are dropped from the queue.

Fig. \ref{7-25} shows the gain of the per packet average delay performance comparison with the beam width varying in the street canyon scenario. This scenario is still unsaturated. When the beam width is wide, the gain is small. The reason can be explained as follows. In the outdoor scenario, wider beam width covers larger area, which makes the beamtracking easier to track the varying location. On the other hand, when the beam width is getting narrow, the gain is significantly getting larger. More narrow beam width makes the beamtracking more difficult to track the varying location. Considering that mmWave wireless network usually has narrow beam, the SAC is more applicable to mmWave wireless network rather than the sub-7GHz band.

The simulation results confirm the performance gain and influencing factors of sensing-assisted communication analyzed above. Tab. \ref{tab:1} summarize the performance evaluation insights from system simulation.

\section{Future Directions of SAC for the Next Generation WLAN}

\begin{figure}[!t]
\centering
\includegraphics[width=3in]{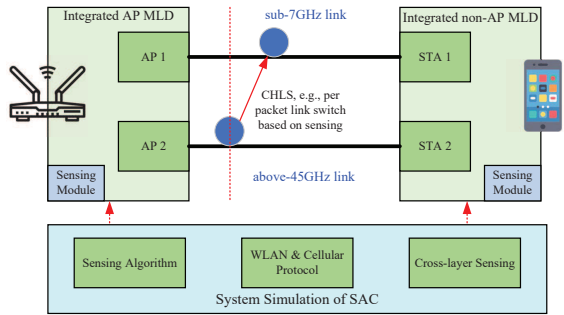}
\caption{Standardization of SAC.}
\label{fig_standardization}
\end{figure}

Now, the ISAC including the SAC is a hot topic for the academia. Meanwhile, the IEEE 802.11bf task group is committed to the standardization of WLAN sensing, which is the first step of the SAC standardization. To deeply excavate the advantages of SAC, the industry and the academia need to deepen the SAC research SAC. As the Fig. \ref{fig_standardization} shown, from our opinion, the following features are suggested to be studied:

\subsection{Cross High-and-low-band Sensing (CHLS)}

The WiFi 7 standardized by IEEE 802.11be introduces multiple link operation (MLO) technology by co-locating multiple STAs working at different sub-7GHz bands into one device named multi-link device (MLD). Gan et al.\cite{InteHighLow} propose to integrate sub-7GHz band and above-45GHz band in one standard. We believe that high frequency band and low frequency band have different channel and transmission properties, leading to diverse performance advantages. Therefore, cross high-and-low-band sensing (CHLS) is a promising way. Each packet can be transmitted or retransmitted dynamically in the high frequency band link or in the low frequency band link according to the surrounding sensing results. Moreover, CHLS is better compatible with MLO specified WiFi 7, which smoothens the standard evolving.

\subsection{Performance Evaluation of Sensing Algorithm}

As this article analyzed, the sensing accuracy and real-time capability deeply affect the performance gain of SAC. This article analyzes the gain, influencing factors, and the performance from the system simulation perspective. In the future, many valuable theoretical and practical sensing algorithms can be modeled and implemented. Therefore, the system simulation will provide more interesting and valuable results, from which we can analyze and obtain more profound influencing factors and influential relationships.

\subsection{Cross-layer Sensing}
Sensing space contains very rich factors such as user preference, traffic characteristics, location and mobility features, and channel information. These factors are affiliated with different network layers. Simply standardize only one factor such as location is far from taking advantage of the SAC. Therefore, the study of cross-layer sensing assisted communication is a promising direction.

\subsection{System Simulation towards Multiple networks}
The next generation WLAN especially the millimeter wave WLAN is the target analysis and simulation scenario of this article. SAC can be used in many wireless network scenarios. Therefore, different kinds of the next generation wireless networks such as 6G and WiFi 8 can be studied and simulated by our designed system level simulation platform. More diverse results can be obtained.

\section{Conclusion}
The SAC is a promising enabler for network intelligence. This article is the first work to analyze the source of performance gain and important influence factors from the system simulation perspective. We carefully evaluate the performance and confirm that the SAC obtains significant performance advantages comparing with IEEE 802.11ay. Several insights on the future directions are discussed. The WLAN has some unique features, such as random access, and ubiquitous interference and collisions. These unique features make WLAN face some more tricky challenges compared with cellular networks, which needs to be solved urgently with the help of SAC. We highlight that although this article focuses on WLAN, some analysis and especially the insights from the system simulation can be extended into other wireless networks.

%\section*{Acknowledgments}
%This work was supported in part by the National Natural Science Foundations of CHINA (Grant No. 61871322, No. 61771392, and No. 61771390), and Science and Technology on Avionics Integration Laboratory and the Aeronautical Science Foundation of China (Grant No. 20185553035 and No. 201955053002).


\bibliographystyle{IEEEtran}
\bibliography{mybibfile}

\newpage

\vfill

\end{document}